\documentclass[useAMS,usenatbib]{mn2e}
\usepackage{graphicx}
\newcommand{\hangpar}{\noindent\hangindent.1in}
\newcommand{\teff}[1]{$T_{\rm eff}$}
\newcommand{\vsini}[1]{$v\cdot\sin(i)$}

\title[Heavy Calcium in CP Stars]{Heavy Calcium in CP Stars\thanks{Based on observations 
obtained at the European Southern
Observatory, Paranal and La Silla, Chile (ESO programmes 076.D-0169(A) and 076.C-0172(A)).}}
\author[C. R. Cowley, S. Hubrig, F. Castelli, J. F. Gonz\'{a}lez,
and B. Wolff ]
{C. R. Cowley${^1}$,
\thanks{E-mail: cowley@umich.edu\,(CC); shubrig@eso.org\,(SH);\newline castelli@ts.astro.it\,(FC);
\,fgonzalez@casleo.gov.ar\,(FG);\newline bwolff@eso.org (BW)}
 S. Hubrig${^2}$, F. Castelli${^3}$,
J. F. Gonz\'{a}lez${^4}$, and B. Wolff${^5}$   \\
$^{1}$Department of Astronomy, University of Michigan,
   Ann Arbor, MI 48109-1042, USA\\
$^{2}$European Southern Observatory, Casilla 19001,
Santiago 19, Chile\\
${^3}$INAF-Osservatorio Astronomico di Trieste, Via G. B. Tiepolo 11,
34131, Trieste, Italy \\
${^4}$Complejo Astron\'{o}mico El Leoncito, Casilla 467, 5400
San Juan, Argentina \\
${^5}$European Southern Observatory, Karl-Schwarzschild-Str. 2,
85748, Garching, Germany
}
\begin{document}

\date{Accepted . Received ; in original form }

\pagerange{\pageref{firstpage}--\pageref{lastpage}} \pubyear{2002}

\maketitle

\label{firstpage}

\begin{abstract}
Large wavelength shifts of infrared triplet lines of
Ca {\sc ii} have been observed in the spectra of HgMn and
magnetic Ap stars.  They have been
attributed to  the heavy calcium isotopes,
including $^{48}$Ca.
One member of the triplet, $\lambda$8542,
had been either unavailable, or of poor quality in
earlier spectra.  The present material shows conclusively
that the stellar $\lambda$8542 shifts are consistent
with an interpretation in terms of $^{48}$Ca.
We  find
no relation between isotopic shifts of the Ca {\sc ii} triplet
lines, and those of Hg {\sc ii} $\lambda$3984.
There is a marginal indication that the shifts
are {\it anticorrelated}
with the surface field strengths of the magnetic stars.
We see sparse evidence for $^{48}$Ca in other chemically
peculiar stars, e.g. Am's, metal-poor stars, or
chemically peculiar red giants.  However, the sample
is still very small, and the
wavelengths of all three triplet lines,
including those in the Sun, show slight positive shifts
with respect to terrestrial positions.

Some profiles of the Ca {\sc ii} infrared triplet in the magnetic
stars show extensive wings beyond a well-defined core.
We can obtain reasonable fits to these profiles
using a stratified calcium abundance similar to that used
by previous workers.  There is no indication that either
the stratification or the Zeeman effect significantly
disturbs the measurement of isotope shifts.

\end{abstract}


\section{Introduction}
Calcium is the only chemical element with two
stable, doubly-magic
isotopes.  The two nuclides, $^{40}$Ca and $^{48}$Ca, are
the lightest and heaviest of calcium's six stable isotopes.
Clayton (2003) describes the cosmochemistry of calcium
with an emphasis on the unusual properties of its isotopes.

The infrared triplet of Ca {\sc ii} arises from $3d$ levels, which
fall between the ground $4s$ and $4p$ terms giving rise to the
H and K resonance lines at $\lambda\lambda$3933 and 3968.
The lines of the triplet have wavelengths
8498.02, 8542.09, and 8662.14 \AA.  The precise values
differ slightly, depending on the source.  We adopt here
the values given in the NIST online Handbook
(Sansonetti \& Martin 2003).

Differences less than
0.01~\AA~\ occur between these values and those obtained
from subtracting energy levels in Sugar and Corliss (1985).
They are unimportant in the present context.

Laboratory work on the infrared
triplet was described by
N\"{o}rtersh\"{a}user, et al. (1998).  They report shifts
of +0.21, +0.20, and +0.20~\AA~\ for $\lambda\lambda$8662,
8542, and 8498 respectively for $^{48}$Ca {\sc ii} relative
to $^{40}$Ca {\sc ii}.  Such displacements are
easily measured on high-resolution astronomical spectra.
The shifts for the infrared triplet are
much larger than corresponding shifts for the $4s-4p$
resonance lines ($\approx$ 0.01~\AA, M\r{a}rtenson-Pendrill
et al. 1992).  The large shifts are due to the
high angular momentum and relatively concentrated charge
density of the $3d$ electron of the lower term of the
triplet.

The wavelengths of the triplet for isotopically pure
calcium are given in Table \ref{tab:Nor}
from N\"{o}rtersh\"{a}user, et al. (see their Table 2)
but converted to Angstrom units.  Note that these authors
give shifts only of $\lambda$8542 for $^{46}$Ca.  They
investigated hyperfine structure for the
odd-A isotope, $^{43}$Ca, but the widths are relatively
small, and are ignored here.

\begin{table}
 \label{tab:Nor}
 \centering
  \caption{Isotopic Wavelengths of Ca {\sc ii} Infrared Triplet
  (N\"{o}rtersh\"{a}user, {\it et al.} 1998)}
  \begin{tabular}{c c c c}
  \hline
   Mass No. &Laboratory $\lambda$[\AA]
& Isotopic $\lambda$[\AA]& Shift[\AA] \\ \hline
42& 8498.018&  8498.074&  0.057 \\
42& 8542.089&  8542.146&  0.057 \\
42& 8662.140&  8662.199&  0.059 \\
  &         &          &        \\
43& 8498.018&  8498.101&  0.083 \\
43& 8542.089&  8542.173&  0.084 \\
43& 8662.140&  8662.226&  0.087 \\
  &         &          &        \\
44& 8498.018&  8498.126&  0.108 \\
44& 8542.089&  8542.198&  0.109 \\
44& 8662.140&  8662.252&  0.113 \\
  &         &          &        \\
46& 8542.089&  8542.247&  0.158 \\
  &         &          &        \\
48& 8498.018&  8498.218&  0.200 \\
48& 8542.089&  8542.291&  0.202 \\
48& 8662.140&  8662.347&  0.208 \\ \hline
\end{tabular}
\end{table}

Castelli and Hubrig (2004) first announced detection of
$^{48}$Ca in several HgMn (CP3) stars.  Cowley and Hubrig (2005)
subsequently found a similar effect in magnetic Ap (CP2) stars.
The shifts appear to vary from those indicating nearly
pure $^{48}$Ca to a virtual terrestrial mix (nearly pure
$^{40}$Ca).  Most stars
show some shifts indicating heavy Ca isotopes,
as we will illustrate in the figures below.
This is true for the Sun also, though the significance
of this observation is yet unclear.

\section{Observations and their Measurement}
\label{sec:obs}

The spectra used in this study have been obtained either with 
the UV-Visual Echelle Spectrograph (UVES) at UT2 of the VLT or with the FEROS echelle
spectrograph on the 2.2\,m telescope at La Silla. 
We used the UVES dichroic standard settings covering the spectral 
range from 3300 to 9300\,\AA{}. 
The slit width was set to $0\farcs{}3$
for the red arm, corresponding to a resolving power of 
$\lambda{}/\Delta{}\lambda{} \approx$~110,000. For the blue arm, we used 
a slit width of $0\farcs{}4$ to achieve a resolving power of 
$\approx$~80,000.
In November 2004,
three new standard Dichroic settings with a central wavelength  at
760\,nm (Dic\#2 346+760, 390+760, 437+760) were implemented.
This reconfiguration made it possible to measure the $\lambda$8542-line,
which had previously been unavailable because of an order gap.

The UVES spectra were reduced by the UVES pipeline Data Reduction
Software (version 2.5), which is an
evolved version of the ECHELLE context of MIDAS.
The S/N ratios of the obtained spectra differ, but
range from about 100 to 400.

The wavelength coverage for the FEROS spectra is from 3530 to
9220\,\AA{}, but often omitting the $\lambda$8542 region.
The nominal resolving
power is 48,000, with a S/N ratio of ca.\ 150-400.
Basic steps of spectroscopic reduction
(bias subtraction, division by a normalized flat field,
extraction of a 1D spectrum and
wavelength calibration) were based on IRAF.

The spectra were mildly Fourier filtered
before the wavelengths were measured at Michigan.
For a few stars, the wavelength positions were
also determined by comparing the observed spectra with
synthetic spectra.
Table A1 gives measurements for all of
the Ca {\sc ii} lines used in this paper.


\section{The Stellar Isotopic Shifts}

Stellar wavelengths often differ slightly from their laboratory
positions.
For atomic spectra, this is only rarely due to isotopic
effects.  Far more often, two atomic lines blend together
so that the measured position corresponds to a weighted
mean of the two laboratory positions.  Generally speaking,
the stronger the line, the less likely its wavelength is
to be perturbed.  This is generally the situation with the
Ca {\sc ii} lines in Multiplet 2, known as the infrared triplet.

Wavelengths of symmetrical, unblended lines may be
measured to an accuracy better than $\pm$ 0.01~\AA\ on
high-resolution spectra of the kind discussed here.
For asymmetrical features, such as those influenced
by blending, differences up to 0.05~\AA\ may occur
for repeated measurements of the same feature; one
measurement might reflect the center of gravity of a
feature, another the flux minimum.

The lines of the calcium infrared triplet are all formed
on the wings of broad Paschen lines: P13, $\lambda$8665.02;
P15, $\lambda$8545.39; and P16, $\lambda$8502.49.
These hydrogen lines influence the depths of the
nearby calcium features but any wavelength perturbations
are below the level of accuracy considered here.
Other nearby atomic lines are usually too weak to
influence the measured wavelengths adopted for the
calcium lines.  Weak blends may be of some influence
when combined with the effect of blending of isotopic
profiles, or especially with Zeeman splitting.

We allow a liberal overall uncertainty
in the measured positions, of about $\pm$ 0.05~\AA.
Typical errors should be somewhat smaller than this,
perhaps $\pm$ 0.03~\AA, as indicated in the figures.

\begin{figure}
\includegraphics[width=55mm,height=84mm,angle=270]{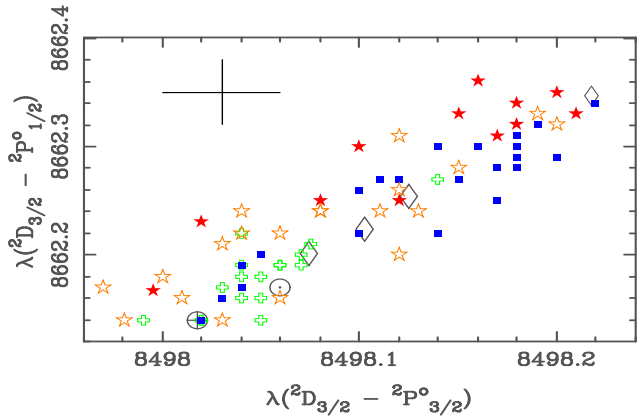}
 \caption{Wavelengths of Ca {\sc ii} $\lambda$8662 are
correlated with
those for $\lambda$8498.
Filled squares are HgMn stars;
filled and open stars are for roAp and noAp stars
respectively.  Open crosses are for miscellaneous types.
Solar wavelengths are indicated by the sun symbol.
Diamonds indicate wavelengths for pure isotopes
of mass 42, 43, 44, and 48, as
given by N\"{o}rtersh\"{a}user, et al. (1998).
Wavelengths from the NIST web site are indicated
by an Earth symbol ($\oplus$); the terrestrial
mix is 97\% $^{40}$Ca.  Error estimates are
indicated in the upper-left portion of the
figure.}
\label{fig:ca9862}
\end{figure}

In Fig.~\ref{fig:ca9862}, all wavelengths listed in
Tab.~A1 for $\lambda\lambda$8498 and 8662
are plotted.  The number of objects previously studied for the
$^{48}$Ca anomaly are substantially increased over previous
work (cf. Castelli \&  Hubrig 2004, Cowley \&
Hubrig 2005).


Fig.~\ref{fig:ca9862} shows that wavelength shifts for
Ca {\sc ii} $\lambda$8498 and $\lambda$8662 are correlated
with one another, as one would expect if they are
isotopic.  It is intriguing that most stars, {\it
including the Sun}, have shifts that could be interpreted
as having a larger admixture of heavier isotopes
than in terrestrial materials.

Chmielewski (2000) noted these shifts for the solar
lines, but the recent study of calcium in late-type
stars (Mashonikina, Korn \& Przybilla 2007) does not mention
them.  This may be due to the explicit inclusion of the
isotopic wavelengths in the later work; unfortunately
these authors do not give figures for the infrared triplet
lines.  Our own calculations including isotopic wavelengths
produce a definite shift of the
profiles to the red, though we still do not get
satisfactory overall fits.

The discrepancy between the solar and laboratory wavelengths
(0.03 to 0.04~ \AA) may arise from the fact that the
laboratory measurements reflect wavelengths made on
optically thin sources.  Small contributions from the
rare isotopes could then be missed.  On the other hand,
the solar features arise from lines that are optically
thick.  This would make the isotope shifts more evident.
Thus, the small shifts are not necessarily due to isotopic
anomalies.

Note that
in a few instances, wavelengths {\it short} of the
terrestrial position, $\lambda$8498.02, were measured.
Similar results were occasionally obtained for the
$\lambda$8542 and $\lambda$8662 lines.
Some of these results are arguably due to measurement error.
Other negative shifts occur for double-lined binaries
(SB2), and are plausibly due to perturbations from the
secondary spectrum.

The case for isotopic shifts is considerably strengthened
by reliable measurements for the third line of the triplet,
$\lambda$8542.  Most of these became available only after
the reconfiguration
of the UVES spectrograph.

\begin{figure}
\includegraphics[width=55mm,height=84mm,angle=270]{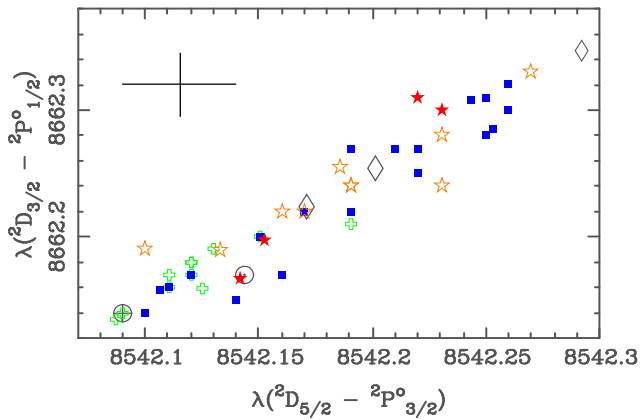}
 \caption{Measured wavelengths Ca {\sc ii} $\lambda$8662 {\it vs.}
those for $\lambda$8542.  The symbols have the same meaning
as in the previous figure.}
\label{fig:ca4262}
\end{figure}

The results are shown in Fig.~\ref{fig:ca4262}.

We can see that the wavelength displacements
of all three of the Ca {\sc ii}
infrared triplet are closely correlated.


\section{Relation to Isotopic Anomalies in Mercury}

Several elements have
isotopic anomalies in CP stars: He, Hg, Pt, and Tl.  We
restrict consideration here to mercury, which has been
investigated in a large number of CP3 (HgMn) stars.
While both Hg and Pt are arguably present also in
magnetic Ap or CP2 stars, their higher
spectral line densities
make conclusions about isotopic compositions difficult.

Bidelman (1962) reported the tentative identification of
$\lambda$3984 in the spectrum of $\kappa$ Cnc as Hg {\sc ii}.
He supplied additional details of his work to
Federer (1962), along with the interpretation that the mean
wavelength variation from star to star
of the Hg {\sc ii} feature
indicated changes in the isotopic composition.  Bidelman
suggested that the $\lambda$3984 wavelength in $\chi$ Lup
indicated nearly pure $^{204}$Hg.

\begin{figure}
\includegraphics[width=55mm,height=84mm,angle=270]{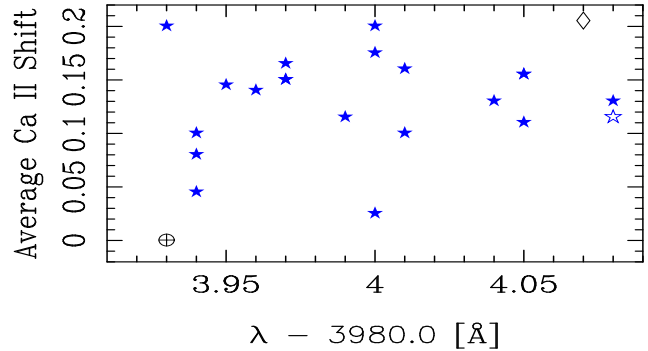}
 \caption{
 Average Ca {\sc ii} shifts ($\lambda$8498 and $\lambda$8662)
 {\it vs.} Hg {\sc ii} $\lambda$3984 wavelength.
 There is no apparent correlation of calcium
 and mercury isotopic anomalies.
 Filled stars are HgMn stars.
 The open star is for Feige 86.  Wavelengths for terrestrial
calcium and mercury are indicated by
an Earth symbol ($\oplus$); the diamond is for pure
$^{48}$Ca and $^{204}$Hg.  We see no indication that isotopic
anomalies are related.}
\label{fig:cahg}
\end{figure}

Fig.~\ref{fig:cahg} gives no indication that the
isotopic anomalies of calcium and mercury are correlated.

\begin{figure}
\includegraphics[width=55mm,height=84mm,angle=270]{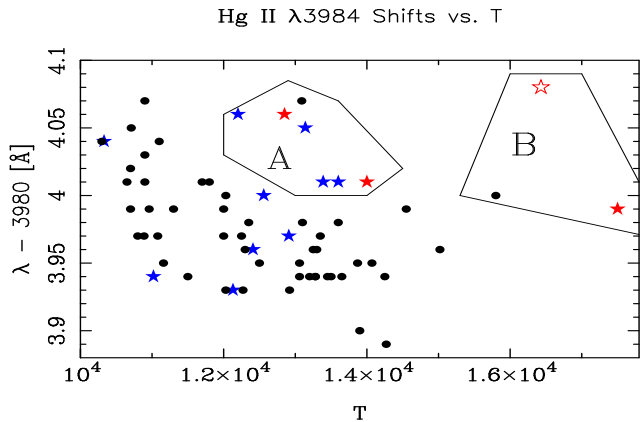}
 \caption{New measurements weaken the temperature
 correlation with isotopic anomalies in mercury.
 Measurements from previous workers are shown as filled
 circles.
 The star-symbols are measurements from ESO spectra.
 The solid star in Region B is 3 Cen A, while
 the open star is Feige 86.  See text for a discussion
 of the objects in Region A.
 }
\label{fig:hgt}
\end{figure}

Preston
(cf. Preston, et al. 1971) and his coworkers noted
a correlation of the isotopic mixture of mercury with
stellar temperature.    Interestingly, newer measurements of
HgMn stars
weaken the temperature-wavelength correlation discussed
by previous workers.  This may be seen from
our Fig.~\ref{fig:hgt}.  The closed circles are
from
Woolf and Lambert (1999),
White, {\it et al.} (1976), or Cowley and
Aikman (1975).  Neither Cowley and Aikman
nor White, {\it et al.} had stars with
temperatures in the range 12,000 to 14,000K
with Hg {\sc ii} wavelengths  $\ge 3984.00$\AA.
But Woolf and Lambert's study already revealed
at least
two ``outliers,'' 46 Aql (12,900K, 3984.073\AA),
and HR 5998 (15,800K, 3984.003\AA).  To these
we add several more.   They fall in two regions
denoted `A' and `B' on Fig.~\ref{fig:hgt}.

Several of the stars in Regions A and B are significantly
different from the ``classical'' HgMn stars,
those studied in the 1970's.  These include
Fiege 86, the field horizontal branch star (cf. Bonifacio,
Castelli \& Hack 2001) and 3 Cen A (Castelli,
Parthasarathy \& Hack 1997) in Region B.
The filled circle in Region B is for HR 5998.  It's
rather broad-lined spectrum resembles
those of the classical HgMn stars.

The filled circle in Region A is 46 Dra,
a ``marginal'' Hg star (Dworetsky 1976).
Region A also contains the unusual HR 6870
(Muthsam and
Cowley 1984) and HR 6000 (Castelli \& Hubrig 2006),
which are
significantly different from classical HgMn stars.
The Hg {\sc ii} features are weak.
We
plot the point for HR 6000 high in Region A, as
originally measured (by CRC), which would indicate nearly
pure $^{204}$Hg.  Castelli and Hubrig's spectrum of HR 6000
shows that lighter isotopes are also
present.  A weighted mean, using the relative abundances
they give would move the point for HR 6000 to
the lower boundary of Region A.
Another
unusual object in Region A is HD 65949 (Cowley, et al.
 2006), whose Hg {\sc ii} line is very strong,
while Mn {\sc ii} is relatively weak.
The sharp-lined spectrum
of $\kappa^2$ Vol is quite similar to that of 46 Aql.
The remaining points in Region A belong to the
double-lined binary $\upsilon^4$Eri.  One
component resembles a classical HgMn star,
while the other has Hg but relatively weak Mn.

Wavelength measurements for Hg II
$\lambda$3984 were mostly taken from
the references cited.  Additional values from
ESO spectra are listed in Table~\ref{tab:newhg},
along with temperatures from the literature,
or based on photometry (Moon 1984).

\begin{table}
 \centering
  \caption{New data used for Fig.~\ref{fig:hgt}}
  \begin{tabular}{l r c c}
  \hline
Name   & HD Number&Temperature&$\lambda-3980.0$ \\ \hline
CD $-$60 1951       &  65949&   13600 &   4.01 \\
CD $-$60 1952       &  65950&   12910 &   3.97 \\
$\kappa^2$ Vol &  71066&   12200 &   4.06 \\
HR 4487& 101189&   11020 &   3.94 \\
3CenA  & 120709&   17500 &   3.99 \\
HR 6000& 144667&   12850 &   4.06 \\
HR 6870& 168733&   14000 &   4.01 \\
HR 7143& 175640&   12130 &   3.93 \\
Feige86&       &   16430 &   4.08 \\  \hline
\end{tabular}
\label{tab:newhg}
\end{table}


\section{Ca {\sc ii} shifts: possible Temperature
Correlations}

\begin{figure}
\includegraphics[width=55mm,height=84mm,angle=270]{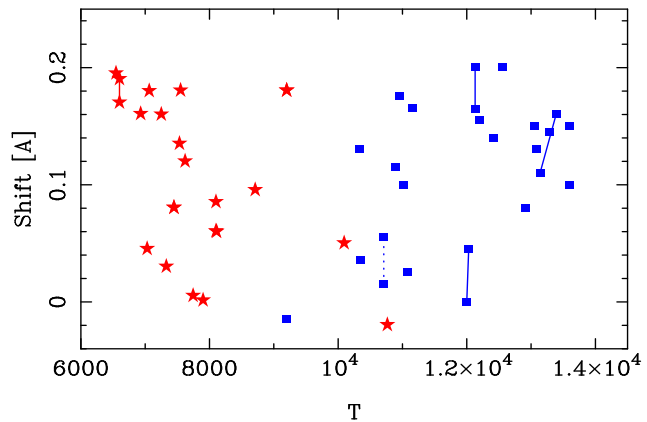}
 \caption{Average shifts of $\lambda\lambda$8498 and 8662
 for magnetic (stars) and HgMn (squares)
 stars are plotted against temperature.  UVES and FEROS
 measures of HD 101065 (far left) and HR 7143 (upper right)
 are connected by a solid line.  Measurements on the
 SB2 stars $\upsilon^4$ Eri (far right), and 74 Aqr
 (lower right) are also connected.  The dashed line
 connects measurements of the minimum (lower point)
 and centroid for the primary of $\chi$ Lup.  See text for references
 to the Hg II wavelength measurements.
 }
\label{fig:sht}
\end{figure}

Fig.~\ref{fig:sht} shows average shifts for the two
Ca {\sc ii} lines for which we have the most measurements,
$\lambda\lambda$8498 and 8662, plotted vs. temperature.
Filled squares are for the HgMn stars.  The overall plot
shows no indication of a correlation of the shifts
with temperature.  If one considers the magnetic
and non-magnetic objects separately, there may be
a weak correlation with temperature but in the
opposite sense for the magnetic and non-magnetic
stars.

Temperatures for the non-magnetic stars are from
sources previously cited.  Most of the temperatures
for the magnetic stars are from Ryabchikova (2005),
or Hubrig, North, and Mathys (2000).  A temperature
of 8100K was used for HD 965, following Bord, Cowley,
Hubrig, and Mathys (2003).

\section{An Inverse Correlation with Magnetic Field Strength?}

\begin{figure}
\includegraphics[width=55mm,height=84mm,angle=270]{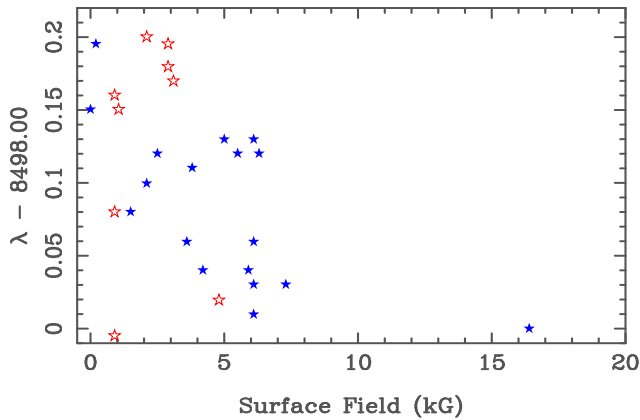}
 \caption{Wavelength shifts of $\lambda$8498 for roAp and
 noAp magnetic stars (open and solid symbols, respectively)
 are plotted vs. the surface magnetic
 fields.  The star with the large field is HD 47103.
 }
\label{fig:mag98}
\end{figure}

Plots of the shifts for both $\lambda\lambda$8498 and 8662
of the magnetic stars suggest that there could be an inverse
correlation of the shifts of these two lines with
surface magnetic field strength.  This is illustrated
for $\lambda$8498 in
Fig.~\ref{fig:mag98}.  A plot for $\lambda$8662
shows a quite similar trend, while one
for $\lambda$8542,
for which we have fewer measurement shows no correlation.
It appears from the plots for $\lambda\lambda$8498, and
8662 that the inverse correlation, if real,
is due to the non-oscillating
magnetic stars (solid stars in Fig.~\ref{fig:mag98}).

We have measurements of one very strong-field star
that definitely contradicts
the overall trend of large shifts with weak fields.
In HD 154708, Hubrig, et al. (2005), measured a mean field
of 24.5 kG.
We measure displacements of +0.14, 0.26, and
0.29 for $\lambda\lambda$8498, 8542, and 8662 respectively.
Though the displacements are all positive,
they are significantly different for one of
the three lines.  Possibly this difference is caused by,
other effects than isotopic shifts--most likely, blending
with highly displaced Zeeman components.
We note with interest, however, the new classification
of HD 154708
as a roAp star (Kurtz, et al. 2006).  This strengthens the
suggestion that it is only the non-oscillating stars that
obey the inverse correlation of field strengths and isotope
shifts.

Clearly
more measurements are necessary to establish this trend.

Magnetic field measurements are from Kurtz, et al. (2006),
Hubrig, et al. (2006), and measured on our spectra
using the magnetically sensitive line Fe II
$\lambda$6149 (cf. Mathys, et al. 1997).  For a few stars
with unresolved $\lambda$6149 and only longitudinal
field measurements available in the literature, we used
the approximate relation
$\langle B_S \rangle \approx 3 \langle B_z \rangle$,
assuming a dipole field.  Error estimates for the fields
vary from star to star, depending on the magnitude of the
Zeeman splitting and the complexity of the profiles.
Likely overall errors in the surface
field values are of the order of a kilogauss.

\section{Profiles of the Ca {\sc ii} Triplet
in the Magnetic Stars}

\subsection{Approximate Calculation of Zeeman Profiles}
We have attempted spectral synthesis of the
Ca {\sc ii} lines in several stars.  For those stars
with wide Zeeman patterns, the spectra were synthesized
using the assumption of a pure longitudinal or a pure
transverse magnetic field.  With these assumptions,
the equations of radiative transfer for the Stokes
parameters become diagonal (cf. del Toro 2003),
so that with the LTE assumption
one may use a standard code.
For the longitudinal
field assumption, separate calculations are made for the
left and right sigma components, and the resultant
flux {\it vs.} wavelength files are averaged with equal
weights.  Similar calculations are made for both
sigma and pi components for the transverse field
assumption.
We approximate the profiles for the general case by
taking a weighted mean for longitudinal and transverse
field orientations.  This scheme is, of course,
inferior to one using a full Stokes transfer, averaging
over the specific intensity, provided the total surface
field orientation is known.  We believe it is a
useful approximation in the absence of detailed knowledge
of the field.

\subsection{The Stellar Profiles}

Even with the simplifications discussed above,
spectral synthesis of the Ca {\sc ii} infrared triplet is
fraught with difficulties.  All three lines are formed
in the wings of the Paschen lines, P13 ($\lambda$8662),
P15 ($\lambda$8542), and P16 ($\lambda$8498).  The
VCS (Vidal, Cooper \& Smith 1973) Stark profiles
do not fit observed lines very well.  This was noted
by Fr\'{e}mat, Houziaux, and Andrillat (1996).

The more recent tabulations of Stehl\'{e} and Hutcheon
(1999), unfortunately, do not extend to large
enough values of $N_e$ (for P13-P16)
to accommodate our models.
We used Lemke's (1997) extended VCS tables, and
leave the problem of fitting the Paschen lines
for another study.

The presence of the Paschen lines, in addition to
significant water-vapor absorption, makes it most
difficult to normalize the observations to a
continuum.  This problem is exacerbated for echelle
observations, where orders must be joined.

In the
figures below, we have adjusted the observations
slightly
(up or down) to optimize the fits with the theory.
This seems permissible, in view of
normalization problems, and the approximate,
working, model atmosphere used.  We have not attempted
to derive accurate abundances from the infrared triplet.
The profile calculations were made to illustrate that
in spite of difficulties with the continuum and the
anomalous line shapes, {\it the cores can be reasonably fit,
and therefore, the measured wavelengths are credible.}

All of the spectra illustrated in this section use
an approximate ATLAS9 (Castelli \& Kurucz 2003)
model with a $T_e$ of 7700K, log($g$) = 4.0,
and Ap-star abundances based on the results
for $\gamma$ Equ (Ryabchikova, et al. 1997).  The
calculations assumed only one dominant calcium
isotope.

\subsection{Stratification}

In the cooler magnetic stars, the lines of the infrared triplet
appear to have very broad wings.  This is illustrated in
Fig.~\ref{fig:theobs42}.

We have been able to account
for the broad wings in $\gamma$ Equ using a stratification
essentially that of Ryabchikova et al. (2002).
For convenience,
we did use an analytical formula based on three
positive parameters,
$a$, $b$, and $d$,
to give the stratification profile.
With $x = \log{\tau_{\rm 5000}}$, we multiply the abundance
by a factor $g(x)$, where

\begin{displaymath}
g(x) = b + (1-b)\left[\frac{1}{2}\sqrt(\pi/a)\pm
\frac{1}{2}{\rm erf}(\sqrt{(a}|x-d|^2)\right].
\end{displaymath}

The negative sign applies for $x \le -d$.
This function is a smoothed step of height
$1-b$, located at depth
$\log{\tau_{\rm 5000} = d}$.
The Gaussian $1/{\rm e}$
width of the step is $1/\sqrt{a}$.
We adjusted the parameters on a trial-and-error
basis rather than attempt an inversion procedure,
as discussed by Kochukhov, et al. (2006).
A recent preprint by Kochukhov (2007) describes a parameterized
approach similar to ours.  In
the present work, we used stratification only for
two stars (see Figures 7 and 8 below).

\subsection{Moderate field strengths: $\gamma$ Equ}

The $\lambda$8498-line was fit for $\gamma$ Equ
by a stratified model using $a = 6.7$, $b = 10^{-4}$,
and $d = 1$.  These parameters put an abundance jump
of nearly 4 dex at $\log{\tau_{5000}} = -1$; the width
of the jump is roughly one unit in
$\log{\tau_{5000}}$.  The stratification is quite similar
to that already used by Ryabchikova, et al. (2002).
We can confirm that the same model accounts quite well
for the stronger Ca {\sc ii} K-line profile.

\begin{figure}
\includegraphics[width=55mm,height=84mm,angle=270]{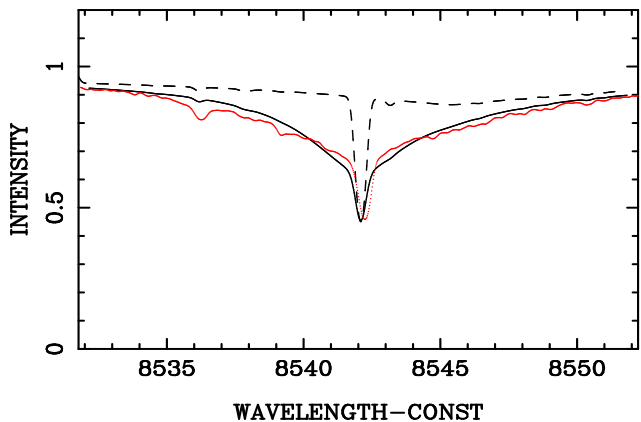}
 \caption{Ca {\sc ii} $\lambda$8542 in $\gamma$ Equ.
 The observed spectrum is dotted.
 The slight
 P15 minimum is seen in the calculations at 8545.4~\AA, but
 the influence of this line is small.
 Note the small
 displacement of the calculated minimum from the observed one,
 which we attribute to an admixture of $^{48}$Ca.
 The solid curve employs the stratification described in the
 text, with Ca/Sum = $3\cdot 10^{-5}$ in the deepest layers.
 This is some 15 times the solar abundance of calcium.
 The dashed line results from no stratification and
 Ca/Sum = $4\cdot 10^{-9}$.  While the core is fit, the
 failure in the wings is catastrophic.
}
\label{fig:theobs42}
\end{figure}

The Zeeman components were not explicitly
calculated, in view of the sharpness of the lines in
this star; an
enhanced microturbulence of 4 km/s was employed to
simulate the magnetic effects.  The results are shown
in Fig.~\ref{fig:theobs42}.

We
synthesized several Ca I near $\lambda$6160, in
$\gamma$ Equ using our default model with and
without stratification.
We find
that we can fit them with an abundance ratio of
calcium to the sum of all elemental abundances,
or Ca/Sum = $6\cdot 10^{-6}$, and
$\xi_t = 4$ km s$^{-1}$, using our model.  This is also
near the value found in the work cited.  For these
relatively weak Ca {\sc i} lines, the stratification was
of minor importance.

\subsection{The Triplet at Higher Field Strengths}

For surface fields of 10 kG or more, the Zeeman patterns
are obvious in stars with minimal rotational broadening.
Components of the $\lambda$8498-line, the weakest of the
three Ca {\sc ii} lines, split into a doublet structure.
The remaining two lines resolve into triplets for
sufficiently high fields.  At our resolution, closely
spaced Zeeman components are not seen individually.

\begin{figure}
 \includegraphics[width=55mm,height=84mm,angle=270]{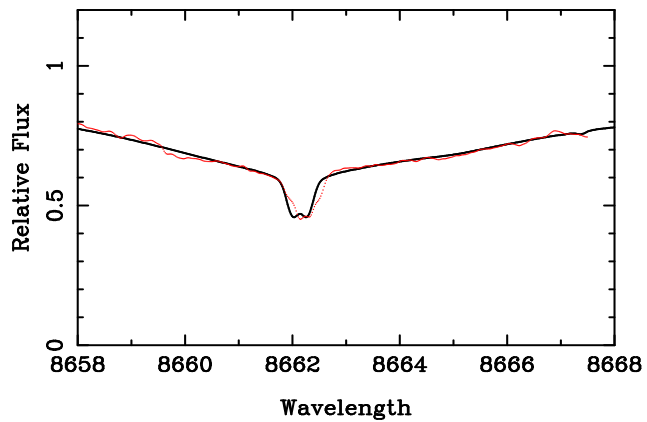}
 \caption{Ca {\sc ii} $\lambda$8662 in HD 51684.  We see again the
 broad wings, suggestive of extreme damping.  The measured
 wavelength shift for this line is 0.11~\AA. The P13 core
is at 8665~\AA.  }
\label{fig:62_51}
\end{figure}

HD 51684 has an surface field of 6.3 kG, according to our
measurement of the splitting of the magnetically sensitive
Fe {\sc ii} line, $\lambda$6149.  However, for the
$\lambda$8662-line, we got a somewhat better fit using
5 kG.
We show
a Zeeman synthesis of the $\lambda$8662 line in
Fig.~\ref{fig:62_51}. where the components are marginally
resolved.  Our best fit (shown) assumed a 3 to 1 ratio of
longitudinal to transverse fields.
The stratification parameters were
$a = 4.0$, $b = 0.001$, and $d = 1.0$.  The figure was
made with
Ca/Sum = $2\cdot 10^{-4}$.

\begin{figure}
\includegraphics[width=55mm,height=84mm,angle=270]{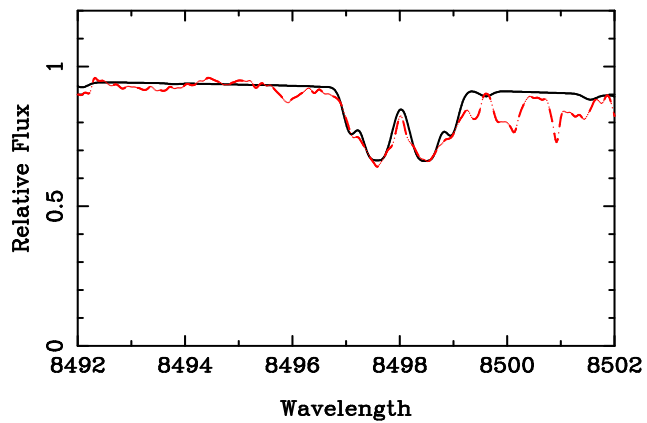}
 \caption{Ca {\sc ii} $\lambda$8498 in HD 47103 (17.5 kG).
 Wavelength measures of lines with
 doublet splitting such as that shown must be of the {\it maximum
 flux} between the two components.  This calculation used
 a longitudinal to transverse field ratio of 2 to 3, and an
 abundance Ca/Sum of $6.5\times 10^{-8}$.  No stratification
 was assumed.
 }
\label{fig:8498_47}
\end{figure}

HD 47103 is one of our stars with the strongest fields.
The $\lambda$8498-line shows a wide doublet structure
(Fig.~\ref{fig:8498_47}).  The broad wings seen in
the previous figures are not evident, and stratification
was not used .

We determined a mean surface field of 16.4kG from
Fe {\sc ii} $\lambda$6149.  However,
the fit shown in Fig.~\ref{fig:8498_47} is slightly
better if we use the surface field of 17.5 kG (cf.
Babel, North \& Queloz 1995).

\begin{figure}
\includegraphics[width=55mm,height=84mm,angle=270]{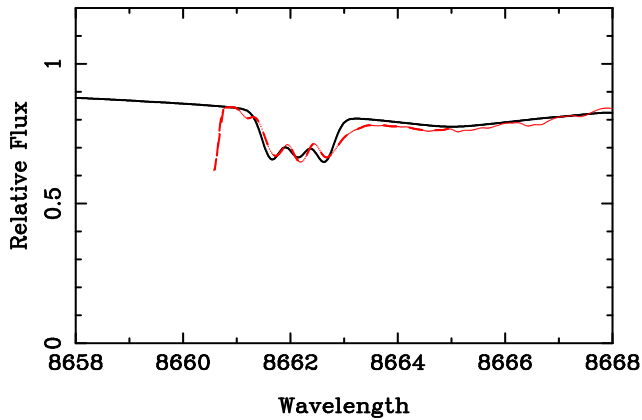}
 \caption{Ca {\sc ii} $\lambda$8662 in HD 47103 (17.5 kG).
 The observational profile on the violet side is
 unavailable because
 of an echelle order gap.
 Optimum fit was obtained with
 slightly different parameters from those used for the
 $\lambda$8498-line.  We used Ca/Sum = $5.5\times 10^{-8}$,
 and a longitudinal to transverse ratio of 1 to 5.
 }
\label{fig:8662_47}
\end{figure}

Fig.~\ref{fig:8662_47} illustrates the triplet structure when
the it is well resolved.  The figure also
shows that this line is very near an order gap in the spectral
coverage.  This gap produces obvious distortions of the
flux.  Because of this, we have been unable to determine
whether this line could be better fit by a stratification
model.
We have arbitrarily raised the observed continuum
to fit the calculations at the violet limit of coverage.
We also tried {\it not} applying such a correction, and
using various stratification models, but these did not
produce improved fits.  Better observations are needed
to resolve this question.  Stratification was not assumed
for either plot (Figs. 9 and 10) for HD 47103.

We find no evidence--outside of our stated errors--that the
measured stellar wavelengths have been affected by proximity
to order gaps.

The optimum fits to the $\lambda$8498 and $\lambda$8662 lines
in Figs.~\ref{fig:8498_47} and ~\ref{fig:8662_47} used
slightly different longitudinal to transverse ratios.
It is unclear whether these differences arise
from the approximate
nature of our Zeeman calculation, or another cause, such as
a partial Paschen-Bach effect (Mathys 1990).

\section{Summary}

This work is an investigation of the overall
interpretation of wavelength shifts in the Ca {\sc ii}
spectrum as isotopic.  Previous investigations had
been based on only two of the three lines of the
infrared triplet.  Following a reconfiguration of the
ESO UVES spectrograph, we were able to measure the
third wavelength, $\lambda$8542, in 41 additional
spectra.  The new measurements showed the expected isotopic
shifts.

We investigated possible correlations of the isotopic
shifts in calcium with temperature, magnetic field
strength, and with isotopic anomalies in mercury.
We find a possible {\it anticorrelation} of
isotope shifts with magnetic field strength, perhaps
limited to non-oscillating Ap stars.  Other correlations,
are marginal at best (Section 6).

New measurements of the Hg {\sc ii} $\lambda$3984
wavelengths showed that the temperature-isotopic anomaly
relation is complicated.
It is extremely puzzling that the Ca-48 anomaly and
the Hg-204 (or just heavy Hg) appear
uncorrelated.  The only current ideas for such anomalies
involve chemical fractionation rather than nuclear processes;
these ideas would lead us to expect some correlation.  The
fact that no correlations are yet evident does not vitiate
the notion of fractionation,
but only means that whatever scenerio eventually emerges
will be complex.

We found a
much weaker overall relation between stellar temperature
and the percentage of heavy
mercury than previous workers.  This is reasonably attributed
to our investigation of a
wider class of objects--a more diverse group--than was
studied in the 1970's (cf. Section 4).

It is too early to conclude that a real, inverse relation
exists between the Ca-48 shifts
and surface field (see Fig.~6).
Perhaps this is fortunate, as at the
moment, we can not suggest a reason for the relation.

We synthesized a number of Ca {\sc ii} lines from
the infrared triplet in several magnetic stars.
In this way we established that valid wavelengths
could be obtained even in the presence of considerable
Zeeman effects, instrumental perturbations, and the
proximity of Paschen lines.

Overall fits of the cores and broad wings of
$\lambda$8498 in $\gamma$ Equ and $\lambda$8542 in
HD 51684 required models including stratification
for two of our stars.  Previous studies have
revealed the need for such models.
These profile anomalies must be considered along with
those for the Balmer lines, and Ca {\sc II}
H and K cores.  They clearly indicate differences
in the high atmospheric structure of magnetic
and normal stars.
Since our principal concern in the present paper
is with the isotopic wavelength shifts,
we leave details of stratified models to the
references cited and future work.

\section*{Acknowledgments}

We thank Dr. G. A. Wade for providing calculations
made with a full Stokes transfer that we could use
to test our Zeeman procedure.  We acknowledge a
useful conversation with Dr. J. D. Landstreet.
This research has made use of the SIMBAD database,
operated at CDS, Strasbourg, France.  We appreciate
the advice from Dr. Jeffrey Fuhr of NIST on
oscillator strengths for the Ca {\sc ii} infrared
triplet.
We acknowledge
use of the Vienna Atomic Database (Kupka,
{\it et al.} 1999), and UVES Paranal Observatory
Project (ESO DDT Program ID266.D-5655, Bagnulo, et al.
2003)

\section*{REFERENCES}


\hangpar Babel, J., North, P., Queloz, D., 1995, A\&A, L5

\hangpar Bagnulo, S., Jehin, E., Ledoux, C., et al., 2003,
ESO Mess., 114, 10



\hangpar Bidelman, W. P., 1962, AJ, 67, 111

\hangpar Bonifacio, P., Castelli, F., Hack, M., 2001,
A\&AS, 110. 441

\hangpar Bord, D. J., Cowley, C. R., Hubrig, S., Mathys, G.
2003, Bull. AAS, 35, 1357

\hangpar Castelli, F. \&Hubrig, S, 2006, http://wwwuser.oat.ts.it/
castelli/hr6000/hr6000.html

\hangpar Castelli, F., Hubrig, S., 2004, A\&A, 421, L1


\hangpar Castelli, F., Kurucz, R. L., 2003, IAU Symposium, 210, 20P

\hangpar Castelli, F., Parthasarathy, M., Hack, M., 1997,
A\&A, 321.254

S

\hangpar Chmielewski, Y., 2000, A\&A, 353, 666

\hangpar Clayton, D., 2003,
{\it Handbook of Isotopes in the Cosmos}, Cambridge, University
Press (see p. 196)

\hangpar Cowley, C. R., Aikman, G. C. L., 1975, PASP, 87, 513

\hangpar Cowley, C. R., Hubrig, S., 2005, A\&A, 432, L21

\hangpar Cowley, C. R., Hubrig, S., Gonz\'{a}lez, J. F.,
Nu\~{n}ez, N., 2006, A\&A, 455, L21

\hangpar del Toro Iniesta, J., C., 2003, {\it Introduction
to Spectropolarimetry}, (Cambridge: University Press),
cf. 7.8 and 9.4.2


\hangpar Dworetsky, M. M., 1976, in IAU Colloquium 32,
ed. W. W. Weiss, H. Jenkner, H. J. Wood, p 553

\hangpar Federer, C. A., 1962, Sky and Telescope, 23, 140

\hangpar Fr\'{e}mat, Y., Houziaux, L., Andrillat, Y., 1996,
MNRAS, 279, 25





Academic Press


\hangpar Hubrig, S., North, P, Mathys, G., 2000, ApJ, 539,
352

\hangpar Hubrig, S., Nesvacil, N., Sch\"{o}ller, M., et al.,
2005, A\&A, 440, L37.

\hangpar Hubrig, S., North, P., Sch\"{o}ller, M., Mathys, G.,
2006, AN, 327, 289


\hangpar Kochukhov, O., 2007, astro-ph/0701084 (Sec. 3.4)

\hangpar Kochukhov, O., Tsymbal, V., Ryabchikova, T., et al.,
A\&A, 2006, 460, 831

\hangpar Kupka, F., Piskunov, N, E., Ryabchikova, T. A.,
Stempels, H. C., Weiss, W. W., 1999, A\&AS, 138, 119

\hangpar Kurtz, D. W., Elkin, V. G., Cuhna, M. S., et al.,
2006, MNRAS, 372, 286





\hangpar Lemke, M. 1997, A\&AS, 122, 285

\hangpar M\r{a}rtensson-Pendrill, A.-M, Ynnerman, A.,
Warston, H., et al., 1992, Phys. Rev. A, 45, 4675

\hangpar Mashonkina, L., Korn, A. J., Przybilla, N., 2007,
A\&A, 461, 261

\hangpar Mathys, G., Hubrig, S., Landstreet, J. D., Lanz, T.,
Manfroid, J. 1997, A\&AS, 123, 353

\hangpar Mathys, G., 1990, A\&A, 232, 151

\hangpar Moon, T. T., 1984, Comm. Univ. London Obs., No. 78

\hangpar Muthsam, H., Cowley, C. R., 1984, ApJ, 130, 348

\hangpar N\"{o}rtersh\"{a}user, W., Blaum, K., Icker, P.,
et al., 1998, Eur. Phys. J. D, 2, 33

\hangpar Preston, G. W., Vaughan, A. H., White, R. E.,
Swings, J-P. 1971, PASP, 83, 607

\hangpar Ryabchikova, T. A., 2005, Pis'ma Astron. Zh., 31,
437 (see also CDS, VIZIE-R J/PAZh/31/437).

\hangpar Ryabchikova, T. A., Adelman, S. J., Weiss, W. W.,
Kuschnig, R., 1997, A\&A, 322, 234

\hangpar Ryabchikova, T., Piskunov, N., Kochukhov, O., {\it et
al.}, 2002, A\&A, 384, 545


\hangpar Sansonetti, J. E., Martin, W. F., 2003, Handbook
of Basic Atomic Spectroscopic Data, http://physics.nist.gov/
PhysRefData/Handbook/index.html


\hangpar Stehl\'{e}, C., Hutcheon, R. 1999, A\&AS, 140, 93

\hangpar Sugar, J., Corliss, C. H., 1985, J. Phys. Chem. Ref.
Data, 14, Suppl. No. 2

\hangpar White, R. E., Vaughan, A. H., Jr., Preston, G. W.,
Swings, J. P., 1976, ApJ, 204, 131

\hangpar Vidal, C. R., Cooper, J., Smith, E. W., 1973, ApJS, 25, 37

\hangpar Woolf, V. M., Lambert, D. L. 1999, ApJ, 521, 414


\appendix

\section{Wavelength Measurements of Ca {\sc ii} Infrared Triplet}

Table A1 lists the program stars and the measured Ca {\sc ii} wavelengths.
The $\lambda$8542-line was unavailable for many of the stars
before reconfiguration of the UVES spectrograph (Section \ref{sec:obs}).
FEROS spectra sometimes included the $\lambda$8542 line.

The stars are listed
in order of their HD numbers within three categories,
Mercury-Manganese and related stars, traditional magnetic
Ap stars, and
miscellaneous types.  The traditional magnetic Ap stars as
used here have strong iron-group and lanthanide spectra.
Clearly some assignments are arguable, for example, HR 6870
to miscellaneous, and HD 101065 to the traditional category.
Oscillating (roAp) and non-oscillating (noAp) stars are indicated.

Entries for HgMn stars and miscellaneous types
with the remark ``CH04'' are from work of
Castelli and Hubrig (2004).
Previously unpublished
measures on FEROS spectra are labeled ``new.''
Entries for magnetic stars and miscellaneous type stars
with the remark ``CH05'' are from work
by Cowley and Hubrig (2005).

A few notes on individual cases follow.

HD 1909 (HR 89): Rather broad lines.  Measurements are
for the primary in both entries.  Lines from the
secondary are difficult to pick out.

HD 32964 (66 Eri): The measurements are for primary and
secondary.

HD 34364 (AR Aur): Measurements for violet($-$) and red
(+) systems of the SB2 were made on different spectra.  The
profiles belonging to the violet system were of poor
quality.

HD 60435: Spectra of this star were of poor quality

HD 65949: The two entries are on from the same FEROS spectrum made
on different days.

HD 93507: The $\lambda$8662 profile was poor.

HD 124740 and HD 141556 ($\chi$ Lup): The minima of all
three profiles for primaries (A) of both stars
show no shift, but all three profiles have shallow slopes
on the red side, indicative of blending.  This cannot be
Ca II of the secondary, because those lines lie to the
violet.

HD 168733:  The same UVES spectrum was measured on two different days.

HD 201601 ($\gamma$ Equ): Measurements on three different exposures
are presented.

\begin{table*}
 \label{tab:A1}
 \centering
  \caption{Stellar Wavelength Measurements}
  \begin{tabular}{r l l r r r l l }
  \hline
\multicolumn{1}{c}{HD}  &
\multicolumn{1}{c}{Other} &
\multicolumn{1}{c}{Sp.\ Type and} &
\multicolumn{3}{c}{Nist Wavelengths} &
\multicolumn{1}{c}{Instrument}&
\multicolumn{1}{c}{Remarks} \\ 
       &
\multicolumn{1}{c}{Designation} &
\multicolumn{1}{c}{Binarity} &
\multicolumn{1}{c}{8498.02} &
\multicolumn{1}{c}{8542.09} & 
\multicolumn{1}{c}{8662.14}&  \\
\hline
\hline
  \multicolumn{8}{c}{Mercury-Manganese and Related Stars} \\ 
\hline
   1909& HR 89                & B9IV, SB2     &  0.11    &            &   0.14  & UVES old &CH04\\
   1909& HR 89                &         &  0.14    &            &   0.14  & FEROS new &     \\
  11753& $\phi$\,Phe          & B9, SB1     &  0.15    &  0.13     &   0.11  & UVES new &       \\
  27376& HR 1347 ($-$) & B9, SB2            &  0.09    &  0.10      & 0.13    & UVES new &              \\
  27376& HR 1347 (+)          &         &  0.16    &  0.17      & 0.16    & UVES new &               \\
  32964& 66~Eri~A      & B9V&  0.01    &  0.05      &  $-$0.04& UVES new &       \\
  32964& 66~Eri~B      & B9V     &  0.10    &  0.13      &   0.13  & UVES new &       \\
  33904& $\mu$~Lep     & B9IV    &  0.08    &  0.08      &   0.08  & UVES new &       \\
  34364&  AR~Aur($-$)& B9V, SB2  & $-$0.09  & $-$0.03    &   0.17  & UVES new & bad profiles\\
  34364&   AR Aur($-$) &         & $-$0.04  & $-$0.03    &   0.05  & UVES new & diff expos   \\
  34364&   AR Aur(+)   &         &  0.05    &  0.06      &   0.03  & UVES new &      \\
  34364&   AR Aur(+)         &         &  0.11    &  0.05      &   0.02  & UVES new & \\
  35548& HR 1800             &B9, SB2  &  0.17    &  0.17      &   0.18  & UVES new &  \\
  65949& CD $-$60~1951        &B8/B9    &  0.12    &            &   0.08  & FEROS new &  \\
  65949& CD $-$60~1951        &B8/B9    &  0.12    &            &   0.07  & FEROS new&\\
  65950& CD $-$60~1952        &B8III    &  0.15    &0.15        &   0.15  & FEROS new &\\
 101189& HR 4487              &B9IV    &  0.09    &$-$0.06     &   0.06  & FEROS old&CH04\\
 124740& CD $-$40~8541         &         &  0.00    & 0.00       &   0.00  & UVES new&asym to red  \\
 141556& $\chi$~Lup           &B9, SB2   &  0.00    &  0.00      &   0.00  & UVES new &asym to red \\
 144667& HR 6000              &         &  0.14    &  0.14      &   0.14  & UVES new &\\
 149121& 28 Her               &B9, SB1   &  0.11    &$-$0.04     &   0.04  & FEROS old&CH04\\
 158704& HR 6520              &B9, SB2  &  0.17    & 0.10       &   0.19  & FEROS old&CH04\\
 165493& HR 6759              &B8, SB1  &  0.20    & 0.05       &   0.05  & FEROS old&CH04\\
 175640& HR 7143              &B9III   &  0.20    &  0.20      &   0.20  & UVES old+FEROS old &CH04 \\
 178065& HR 7245              &B9, SB1   &  0.16    &  0.16      &   0.17  & UVES new &       \\
 186122& 46~Aql               &B9III    &  0.12    &  0.12      &   0.12  & UVES new &       \\
 216494& 74~Aqr A     &B8 IV/V, SB2 &  0.03    &  0.06      &   0.06  & UVES new &    \\
 216494& 74~Aqr B     &B8        &0.00    &  0.01      &   0.00  & UVES new     \\ 
 221507& HR 3987              &B9.5 IV  &  0.11    &            &   0.11  & UVES old&CH04 \\
 Feige~86& BD $+$30~2431   &HZB star             &  0.12    &  0.12      & 0.12    & UVES new& \\
\hline

\multicolumn{8}{c}{Typical Magnetic Ap or CP2 Stars} \\  
\hline
    965& BD $-$00~21 & noAp  &  0.02    &           &  0.10   &  UVES old   &CH05     \\
    965&            &noAp   &  0.02    &  0.07     &  0.08   &  UVES new  &      \\
   2453&  BD $+$31~59& noAp  &  0.09    &  0.10     &  0.10   &  UVES new  &               \\
  15144& HR 710   & noAp    & 0.08     &  0.11     & 0.09    &  UVES new  &     \\
  24712&  HR 1217 & roAp    &  0.15    &           &  0.17   &  UVES old  &CH05    \\
  47103&  BD $+$20~1508 & noAp & $-$0.02  &           &  0.04   &  UVES old &CH05     \\
  50169& BD $-$01~1414  & noAp&  0.01    &           &  0.07   &  UVES old  &CH05    \\
  51684& CD $-$40~2796 & noAp            &  0.10    &  0.10     &  0.11   &  UVES new&             \\
  52696& BD $-$19~1651     & noAp          &  0.11    &  0.08     &  0.08   &  UVES new  &               \\
  60435& CD $-$57~1762&roA             &  0.06    &           &  0.11   &  UVES old& CH05\\
  75445& CD $-$38~4907 & noAp            &  0.10    &           &  0.17   &  UVES old  & CH05     \\
  93507& CD $-$67~955  & noAp           &  0.01    &           &  0.00   &  UVES old &CH05\\
  94660& HR 4263& noAp      & $-$0.04  &           &  0.00   &  UVES old  &  CH05     \\
  94660&       & &$-$0.04  & $-$0.05   & $-$0.01 & UVES new   &             \\
 101065& Przybylski's&roAp &  0.16    &           &  0.18   & FEROS old      &CH05    \\
 101065&  &roAp   & 0.19&          &  0.19   & UVES old   &CH05    \\
 116114& BD $-$17 3829 &roAp            &  0.04    &           &  0.02   & UVES old   &CH05      \\
 122970& BD $+$06 2827 &roAp            &  0.13    &           &  0.19   & UVES old   &CH05     \\
 125248& CS~Vir  &noAp     &  0.06    &           &  0.10   & UVES old& CH05, poor\\
 128898&$\alpha$ Cir&roAp  &$-$0.02   &           &  0.03   & FEROS old & CH04\\
 133792& HR 5623 &noAp     &  0.18    &           &  0.18   & UVES old   &CH05       \\
 134214& BD $-$13 4081&roAp               &  0.16    &           &  0.20   & UVES old   &CH05   \\
\hline
\end{tabular}
\end{table*}

\begin{table*}
 \label{tab:A2}
 \centering
  \contcaption{Stellar Wavelength Measurements}
  \begin{tabular}{r l l r r r l l }
  \hline
\multicolumn{1}{c}{HD}  &
\multicolumn{1}{c}{Other} &
\multicolumn{1}{c}{Sp.\ Type and} &
\multicolumn{3}{c}{Nist Wavelengths} &
\multicolumn{1}{c}{Instrument}&
\multicolumn{1}{c}{Remarks} \\ 
       &
\multicolumn{1}{c}{Designation} &
\multicolumn{1}{c}{Binarity} &
\multicolumn{1}{c}{8498.02} &
\multicolumn{1}{c}{8542.09} & 
\multicolumn{1}{c}{8662.14}&  \\
  \hline
  \hline
\multicolumn{8}{c}{Typical Magnetic Ap or CP2 Stars, {\sl continued}} \\  
\hline
 137909&$\beta$ CrB&roAp    &  0.10    &           &  0.06   & UVES old   &CH05    \\
 137949& 33~Lib  &roAp     &  0.00    &           &  0.09   & UVES old   &CH05   \\
 176232& 10~Aql   &roAp    &  0.14    &           &  0.22   & UVES old   &CH05  \\
 187474& HR 7552  &noAp     &  0.02    &           &  0.08   & UVES old   &CH05     \\
 188041& HR 7575    &noAp &0.04 &           &  0.08   & UVES old   &CH05   \\
 201601&$\gamma$ Equ&roAp &0.08 &           &  0.16   & UVES old& CH05    \\
 201601&            &roAp &0.09 &  0.13     &  0.17   & UVES new & 1st file\\
 201601&            &roAp &0.08 &  0.14     &  0.16   & UVES new & 2nd file\\
 216018& BD $-$12~6357&noAp               & $-$0.01  &           &  0.02   &UVES old & CH05    \\
 217522& CD $-$45~14901 &roAp              &  0.18    &           &  0.21   &UVES old & CH05  \\
\hline

\multicolumn{8}{c}{Misc. Types} \\
\hline
       &              &          &            &         &                 \\
    739&$\theta$~Scl&F4 V&  0.05    &            & 0.05    &UVES old &CH05    \\
  27411& HR 1353 &A3m&  0.02    &            & 0.08    &UVES old& CH05    \\
  28187& CD $-$35~1710&G3 IV/V&  0.04    &  0.04      & 0.05    &UVES new & \\
  84461&$o$~Vel&A0 IV&  0.00    &            &  0.00       &UVES old &CH04 \\
  85503&$\mu$ Leo&K2 III&       0.05    &            & 0.06    &UVES new &\\
  91375& HR 4138 &A1 V&       0.00    &  0.00      & 0.00    &UVES new &       \\
  91793& U Ant &N&         0.03    &            & 0.00    &UVES old &CH05           \\
 104978& $o$~Vir&G8 III Ba II&        0.06    &            & 0.07    &UVES new &\\
 124897& Arcturus&K1.5 III&       0.00    &            & 0.00    &UVES old &CH05   \\
 142860&$\gamma$ Ser&F6 IV/V  &  0.03    &            & 0.02    &UVES new &     \\
 146836& HR 6077&F6 III      &  0.04    &            & 0.05    &UVES old &CH05      \\
 168733& HR 6870&peculiar      & $-$0.05  & $-$0.03    & 0.03    &UVES new&    \\
 168733& HR 6870&peculiar      & $-$0.03  & $-$0.02    & 0.01    &UVES new& remeasure   \\
 188136& CD $-$79~790 &Fm $\delta$ Del           &  0.02    &            & 0.05    &UVES new &\\
 193432& $\nu$ Cap &B9  &  0.02    &  0.02      & 0.02    & UVES new   &     \\
 196426& HR7878 & B9.5 V      &  0.06    &$-$0.1      & 0.03    &FEROS old &CH04 \\
 209459& 21 Peg  &B8 III      &  0.02    &  0.03      & 0.04    &UVES new  &      \\
 210027& $\iota$ Peg & F5 V     &  0.01    &  0.03      & 0.03    &UVES new   &        \\
 214994& o~Peg & A1IV       &  0.03    &  0.03      & 0.04    &UVES new   &     \\
\hline
\end{tabular}
\end{table*}

\label{lastpage}

\end{document}